\documentclass[final, 12pt, times, onecolumn,pdflatex,sn-mathphys-num]{elsarticle}

\usepackage[margin=1in]{geometry} % Adjust margins as needed, e.g., left=1.2in, right=1.2in

\usepackage{setspace}
\usepackage{xstring}
\newcommand{\eqrefs}[1]{(\StrSubstitute{#1}{ }{,})} % Custom command kept
\usepackage{graphicx}%
\usepackage{multirow}%
\usepackage{amsmath,amssymb,amsfonts}% % Combined math packages
\usepackage{amsthm}%
\usepackage{mathrsfs}%
\usepackage[title]{appendix}%
\usepackage{xcolor}%
\usepackage{textcomp}%
\usepackage{manyfoot}%
\usepackage{booktabs}%
\usepackage{algorithm}%
\usepackage{algorithmicx}%
\usepackage{algpseudocode}%
\usepackage{listings}%
\usepackage{float} % Needed for [H] figure placement specifier
\usepackage{enumitem}

\usepackage{hyperref}
\usepackage{cleveref}
\usepackage{natbib} % Often required/recommended by elsarticle
\hypersetup{
    colorlinks,
    citecolor=black, % Keep colors black for consistency or change for reviewer ease (e.g., blue)
    filecolor=black,
    linkcolor=black, % Change to e.g., blue if helpful for reviewers
    urlcolor=black   % Change to e.g., blue if helpful for reviewers
}

\usepackage{tabularx}

\begin{document}
%\tableofcontents % Keep commented out for now

\begin{frontmatter} % Added frontmatter environment for elsarticle

\title{A Novel Data Augmentation Strategy for Robust Deep Learning Classification of Biomedical Time-Series Data: Application to ECG and EEG Analysis}

\author[inst1]{Mohammed Guhdar} % Simplified author format
\ead{mohammed.guhdar@uoz.edu.krd} % using \ead for email

\author[inst1,inst3]{Ramadhan J. Mstafa} % Simplified author format
\ead{ramadhan.mstafa@uoz.edu.krd} % using \ead for email

\author[inst1,inst2,inst3]{Abdulhakeem O. Mohammed} % Simplified author format
\ead{a.mohammed@uoz.edu.krd} % using \ead for email

\affiliation[inst1]{organization={Computer Science Department, College of Science, University of Zakho},
%            addressline={},
            city={Zakho},
%            postcode={42002},
%            state={},
            country={Kurdistan Region, Iraq}}
\affiliation[inst2]{organization={Department of Computer Science and Information Technology, The American University of Kurdistan},
            country={Kurdistan Region, Iraq}}
\affiliation[inst3]{organization={PRIME Lab, Scientific Research Center, University of Zakho},
%            addressline={},
            city={Zakho},
%            postcode={42002},
%            state={},
            country={Kurdistan Region, Iraq}}

\begin{abstract}

The increasing need for accurate and unified analysis of diverse biological signals, such as electrocardiogram (ECG) and electroencephalogram (EEG), is paramount for comprehensive patient assessment, particularly in scenarios involving synchronous monitoring. Despite advances in multi-sensor fusion systems, there remains a critical gap in developing unified architectures that can effectively process and extract meaningful features from fundamentally different physiological signals—a crucial prerequisite for any robust biomedical fusion framework. A significant challenge in this domain is also the inherent class imbalance prevalent in many biomedical datasets, often leading to biased performance in traditional methods. This study addresses these critical needs by proposing a novel and unified deep learning framework capable of achieving state-of-the-art performance across these different signal types. Our methodology integrates a ResNet-based convolutional neural network with an attention mechanism, enhanced by an innovative data augmentation strategy involving the time-domain concatenation of multiple augmented variants of each signal, generating more complex and feature-rich representations. Unlike any research before this one, we tended to make signal representations more complex in a scientific way to achieve future-reaching capabilities, which resulted in the best predictions compared to the state of the art. Preprocessing steps included wavelet denoising, baseline removal, and standardization. Class imbalance was effectively managed through the combined use of this advanced data augmentation and the Focal Loss function. To ensure robust generalization and prevent overfitting, we employed several regularization techniques during training. The proposed architecture was rigorously evaluated on three benchmark datasets: the UCI Seizure EEG dataset, the MIT-BIH Arrhythmia dataset, and the PTB Diagnostic ECG Database(using a single channel). The results demonstrate state-of-the-art performance, with achieved accuracies of \textbf{99.96\%}, \textbf{99.78\%}, and \textbf{100\%}, respectively. These achievements underscore the robustness and generalizability of our approach across diverse biomedical signal types and datasets, highlighting its capacity to effectively extract discriminative features from varying signal morphologies and clinical contexts. Furthermore, calculations indicate that proposed architecture requires approximately 130 MB of memory and can process each sample in around 10 milliseconds, suggesting its potential suitability for deployment on low-end or wearable devices. These findings establish a powerful, unified framework for automated and robust analysis of diverse biomedical time-series data, effectively addressing class imbalance.
\end{abstract}


\end{frontmatter} % End frontmatter

\section{Introduction}
The escalating global burden of cardiac and neurological disorders presents a significant challenge to healthcare systems worldwide, necessitating advancements in diagnostic methodologies \cite{gogan2025cardiac, brito2025chikungunya}. Electrocardiograms (ECGs) and electroencephalograms (EEGs) serve as critical diagnostic modalities for these conditions.  However, traditional interpretation of these complex biomedical signals, reliant on visual assessment by clinicians, is inherently subjective, labor-intensive, and may lack the sensitivity required to detect subtle pathological indicators.  This context underscores the pressing need for automated, robust, and accurate analytical tools capable of processing and interpreting biomedical time-series data with enhanced precision and efficiency.

Deep learning methodologies have emerged as a transformative force in diverse domains, and their application to biomedical signal processing holds immense promise \cite{li2024continual, kolhar2024deep}. These techniques offer the potential to overcome the limitations of conventional analytical approaches, enabling the extraction of intricate patterns and features from complex datasets that may be imperceptible to the human eye.  Consequently, the development of sophisticated deep learning architectures for automated ECG and EEG signal classification is rapidly gaining momentum, driven by the prospect of improved diagnostic accuracy and streamlined clinical workflows.

Despite the advancements in deep learning, significant challenges remain in the context of biomedical signal analysis.  ECG and EEG signals are inherently complex, exhibiting substantial inter- and intra-subject variability, and are often contaminated by noise and artifacts \cite{miljkovic2024electrogastrogram, bennia2024comparative, zhang2025jassnet}. Furthermore, clinically relevant datasets frequently exhibit class imbalance, where pathological conditions are underrepresented compared to normal physiological states.  Addressing these inherent complexities necessitates innovative approaches encompassing robust preprocessing strategies, architectural designs resilient to data perturbations, and techniques to effectively manage skewed class distributions.

To address these challenges, this paper introduces a novel deep learning framework for ECG and EEG signal classification.  The proposed architecture is predicated on a ResNet-based convolutional neural network, strategically enhanced with an attention mechanism \cite{li2018rest, niu2021review}.  The integration of an attention mechanism is specifically designed to enable the model to focus on salient temporal segments within the signals, thereby improving feature extraction and discriminative capacity, particularly in the presence of noisy or artifact-laden data. This architectural enhancement represents a key contribution to the field.

Another significant contribution of this work lies in the development and implementation of a novel data augmentation strategy.  We introduce a time-domain concatenation approach, generating augmented signal variants through the application of time warping, cutout, and amplitude jitter.  By concatenating these augmented variants, the model is implicitly trained to achieve invariance to these common signal distortions, thereby enhancing its robustness and generalization capabilities across diverse datasets and clinical settings. This augmentation strategy offers a novel approach to improve model resilience.

Furthermore, the proposed framework incorporates a comprehensive preprocessing pipeline to ensure signal fidelity and mitigate the impact of noise and artifacts. This pipeline includes wavelet denoising, baseline removal, and standardization techniques \cite{jin2024novel, dayananda2024pre}.  Coupled with the utilization of Focal Loss to effectively address class imbalance, these methodological choices collectively contribute to the robustness and clinical applicability of the proposed approach.

The efficacy of this framework is rigorously validated through extensive experimentation on three benchmark datasets: the UCI Seizure EEG dataset, the MIT-BIH Arrhythmia dataset, and the PTB Diagnostic ECG Database \cite{andrzejak2001indications_UCI_EEG_DATASET, moody1990bih, bousseljot2004ptb}. Across these diverse datasets, underscore the robustness and generalizability of the proposed methodology. The results highlight the potential of the integrated approach for accurate and reliable automated analysis of biomedical time-series data, paving the way for improved diagnostic tools and enhanced clinical decision-making. The subsequent sections will provide a detailed exposition of the methodology, experimental design, results, and a comprehensive discussion of the broader implications of these findings.

\section{Related work}
The burgeoning field of automated biomedical signal analysis has witnessed a paradigm shift with the advent of deep learning methodologies. Traditional signal processing techniques, while valuable in feature engineering and initial analysis, often necessitate extensive domain expertise and may struggle to capture the complex non-linear relationships inherent in physiological data \cite{wasimuddin2020stages}.  In contrast, deep learning architectures, particularly convolutional neural networks (CNNs) and recurrent neural networks (RNNs), possess an inherent capacity for automated feature extraction directly from raw or minimally preprocessed signals, thereby offering a potentially more efficient and robust approach to classification tasks across diverse biomedical domains, including ECG and EEG analysis \cite{petmezas2022state, dash2024review}

Recent research has focused on classifying ECG signals for arrhythmia detection using advanced deep learning techniques. For example,\cite{singh2024novel} investigated various classification methods, proposing a Recurrent Convolutional Neural Network (RCNN) approach. Evaluated on the PTB Diagnostic ECG and MIT-BIH Arrhythmias Databases, their RCNN model, enhanced with Grey Wolf Optimization (GWO), achieved a notable accuracy of 98\%, outperforming traditional machine learning algorithms such as Decision Trees, K-Nearest Neighbors, Random Forests, Support Vector Machines, and Logistic Regression. This study highlights the efficacy of recurrent convolutional architectures for achieving high performance in automated ECG classification.

ResNet architectures, with their residual connections, have proven effective for deep CNNs in biomedical signal analysis. Their depth aids in capturing complex temporal dynamics in ECG and EEG signals For instance, \cite{liu2024art} proposed ART-Net, an attention-based hybrid ResNet-Transformer for 12-lead ECG classification on the CPSC2018 (The China Physiological Signal Challenge 2018) dataset, achieving  85.27\% of accuracy, 86.01\% of sensitivity and 85.59\% of specificity, respectively. This demonstrates the ongoing development of ResNet-based models, often incorporating attention mechanisms, to advance ECG analysis.

Attention mechanisms improve deep learning for biomedical signals by focusing on key temporal segments \cite{niu2021review}. In EEG-based motor imagery, \cite{li2024effective} developed a hybrid Convolutional Neural Network and Bidirectional Long Short-Term Memory (CBLSTM) model enhanced with an attention mechanism. This architecture, designed to extract both spatial and temporal features, achieved an average accuracy of 98.40\% on the PhysioNet dataset for decoding motor imagery tasks, demonstrating the significant advantage of incorporating attention for enhanced EEG signal recognition.

The limited availability of large, balanced, and labeled datasets is a key challenge in deep learning for biomedical signal analysis, making data augmentation indispensable for improved generalization. For example, \cite{lim2025specialized} introduced an optimized data augmentation technique for ECG data based on precordial lead angles, demonstrating improved diagnostic accuracy for various cardiac conditions with enhanced efficiency. Similarly, in pulse diagnosis, \cite{fan2025gadm} proposed a novel Generative Adversarial Diffusion Model (GADM) for data augmentation. This method yielded faster and higher-quality samples, leading to a significant boost in hypertension classification accuracy, from 88.28\% to 94.33\%. These studies highlight the crucial role of effective data augmentation in overcoming data limitations and enhancing deep learning performance across different biomedical signal modalities.

Furthermore, addressing class imbalance, a common issue in clinical biomedical datasets where pathological conditions are less frequent, is crucial for accurate classification. Standard loss functions can be biased towards the majority class. Various machine learning techniques address this, as outlined in a survey by \cite{altalhan2025imbalanced}, including data and algorithm-level methods. Among these, Focal Loss has shown promise by focusing learning on hard-to-classify instances. For instance, \cite{al2019densefocal} utilized Focal Loss within a Dense Convolutional Network (DCN) for ECG classification on the imbalanced MIT-BIH database, achieving significant accuracy improvements, demonstrating its effectiveness in handling class imbalance in ECG analysis.

While oversampling is a common approach to address class imbalance, traditional methods relying on linear interpolation for generating synthetic minority samples may fall short in capturing the complexity of intricate data feature spaces, as suggested by \cite{liu2025smote}.

While prior research has demonstrated significant progress in biomedical signal analysis, several limitations motivate the current work. Traditional signal processing methods \cite{wasimuddin2020stages} often require extensive manual feature engineering and may not effectively capture complex patterns compared to deep learning approaches \cite{petmezas2022state, dash2024review}. Although studies like \cite{singh2024novel} achieved high accuracy in ECG classification using RCNN, their findings might be specific to the datasets employed and for biomedical even above 98\% accuracy is required. Similarly, while ART-Net \cite{liu2024art} showed promise for 12-lead ECG, its performance on other ECG types or different datasets remains to be seen. Research on EEG motor imagery using CBLSTM with attention \cite{li2024effective} achieved high accuracy but focused on a specific EEG task. Furthermore, existing data augmentation techniques often target specific signal modalities like ECG \cite{lim2025specialized} or pulse \cite{fan2025gadm}, and while Focal Loss \cite{al2019densefocal} addresses class imbalance in ECG, the broader applicability across different biomedical signals and architectures warrants further investigation. Finally, the limitations of traditional oversampling methods in handling complex data \cite{liu2025smote} necessitate the exploration of more robust data augmentation strategies. These identified weaknesses in the current state-of-the-art highlight the need for a versatile and effective deep learning framework capable of achieving high accuracy across diverse biomedical signals and datasets, which this research aims to address through its novel integrated approach.

Building upon the identified limitations in existing research, this work introduces a novel and highly effective data augmentation strategy. By generating a diverse set of augmented signal variants—including the original signal, noisy, scaled, and shifted versions, alongside advanced transformations such as time warping, cutout, and amplitude jitter—our methodology creates a richer and more comprehensive training dataset. Notably, a key innovation is the time-domain concatenation of these augmented signal variants, resulting in a more complex and feature-rich input signal for each training instance. This advanced data augmentation, in conjunction with our proposed deep learning architecture featuring a ResNet backbone and an attention mechanism, enabled the achievement of state-of-the-art classification accuracy across three distinct and challenging benchmark datasets: the UCI Seizure EEG dataset, the MIT-BIH Arrhythmia dataset, and the PTB Diagnostic ECG Database—all without the need for explicit oversampling techniques. This superior performance across datasets with varying signal modalities (EEG and ECG) and diverse clinical objectives, achieved with a computationally manageable architecture suitable for potential deployment on resource-constrained devices, underscores the efficacy of our combined approach. It presents a compelling and unified framework for enhancing model robustness and generalization in the analysis of imbalanced biomedical time-series data, offering a significant advancement over traditional methodologies.

\section{Materials and Methods}

This section delineates the datasets employed for evaluating the proposed methodology, along with the preprocessing steps and data augmentation applied to the raw signals and the feature scaling techniques utilized prior to model training.  A rigorous description of these procedures is essential for ensuring reproducibility and facilitating comparative analysis with other research in the field of biomedical signal processing.

\subsection{Dataset}

The efficacy of the proposed deep learning framework was rigorously assessed using three publicly available benchmark datasets, each representing a distinct biomedical signal modality and clinical application. These datasets are widely recognized and frequently employed in the evaluation of automated ECG and EEG classification algorithms, thereby providing a robust foundation for comparative performance analysis.

\subsubsection{UCI Seizure EEG Dataset}

For the evaluation of EEG signal classification, we utilized the UCI Seizure EEG dataset \cite{andrzejak2001indications_UCI_EEG_DATASET}. The original dataset comprised EEG recordings from individuals diagnosed with epilepsy, designed for seizure activity detection. We processed this data by dividing the recordings into numerous segments, with each segment lasting 1 second and containing 178 data points sampled at 178 Hz. The dataset is meticulously labeled, categorizing each segment as either indicative of seizure activity or representing non-seizure background brain activity.

\begin{table}[h]
    \centering
    \caption{UCI Seizure EEG Dataset Characteristics}
    \label{tab:uci_characteristics}
    \begin{tabularx}{\linewidth}{X X}
        \toprule
        \textbf{Characteristic} & \textbf{Description} \\
        \midrule
        Number of Records & 500 \\
        Sampling Frequency & 178 Hz \\
        Resolution & Not explicitly specified \\
        Number of Channels & 178 (Data points/features per 1-second segment) \\
        Signal Duration & 1 second (per segment) \\
        Number of Classes & 2 (Non-Seizure/Seizure) \\
        Total Segments & 11,500 \\
        \bottomrule
    \end{tabularx}
\end{table}

\subsubsection{MIT-BIH Arrhythmia Dataset}

To evaluate the performance on ECG signal classification, we employed the MIT-BIH Arrhythmia dataset \cite{moody1990bih}, a widely recognized and extensively utilized resource for the development and validation of arrhythmia detection algorithms. This dataset contains a collection of two-channel ECG recordings, sampled at 360 Hz, obtained from a diverse cohort of patients experiencing various cardiac arrhythmias. The signals are meticulously annotated by expert cardiologists, providing beat-by-beat labels for a wide spectrum of arrhythmia classes, as well as normal heartbeats. The complexity, imbalanced data instances and diversity of arrhythmia morphologies within this dataset present a significant challenge for automated classification systems, making it a valuable benchmark for evaluating the robustness and generalizability of ECG arrhythmia detection methods.

\begin{table}[h]
\centering
\caption{MIT-BIH Arrhythmia Database Characteristics}
\label{tab:mitbih_characteristics}
\begin{tabular}{l|l}
\hline
\textbf{Characteristic} & \textbf{Description} \\
\hline
Number of Records & 48 \\
Recording Duration & 30 minutes each \\
Sampling Frequency & 360 Hz \\
Resolution & 11-bit \\
Voltage Range & 10 mV \\
Number of Channels & 2 (Modified Lead II and V1) \\
Total Segments & 21,890 \\
Number of Classes & 5 \\
\hline
\end{tabular}
\end{table}

\subsubsection{PTB Diagnostic ECG Database}

Complementing the arrhythmia dataset, we further employed the PTB Diagnostic ECG Database \cite{bousseljot2004ptb} to assess the model's capability in classifying a broader spectrum of cardiac pathologies. This database includes 12-lead ECG recordings, sampled at 1000 Hz, from a diverse patient population encompassing individuals with various cardiac conditions, including myocardial infarction, heart failure, and valvular heart disease, as well as healthy control subjects.  The database provides diagnostic labels at the patient level, enabling the evaluation of the system's ability to discriminate between different diagnostic categories based on multi-lead ECG data. The high sampling rate and multi-lead nature of this dataset present a different set of challenges compared to the MIT-BIH dataset, offering a complementary evaluation perspective.

\begin{table}[h]
\centering
\caption{PTB Diagnostic ECG Database Characteristics}
\label{tab:ptb_characteristics}
\begin{tabular}{l|l}
\hline
\textbf{Characteristic} & \textbf{Description} \\
\hline
Number of Records & 549 \\
Number of Subjects & 290 \\
Sampling Frequency & 1000 Hz \\
Resolution & 16-bit \\
Number of Channels & 12 standard + 3 Frank leads \\
Signal Duration & Variable (~30 seconds) \\
Number of Classes & 2 (Normal/Abnormal) \\
Total Segments & 14,552 \\
\hline
\end{tabular}
\end{table}

\subsection{Data Preprocessing}

This section details the preprocessing steps applied to the raw biomedical signals prior to their use in the deep learning model. These steps include wavelet denoising, baseline removal, and standardization, with the aim of enhancing signal quality and preparing the data for effective feature extraction.

\subsubsection{Wavelet Denoising}
The primary goal of wavelet denoising is to remove high-frequency noise components that are often present in biomedical recordings.\\ \\
\textit{Steps:}
\begin{itemize}
    \item \textbf{Decomposition:} The input signal, $s[n]$, is decomposed using the Discrete Wavelet Transform (DWT) with the Daubechies 4 ('db4') wavelet at level 4. This decomposes the signal into approximation coefficients ($cA_4$) representing the low-frequency components and detail coefficients ($cD_1, cD_2, cD_3, cD_4$) representing the high-frequency components. 
    %The decomposition is performed using the \texttt{pywt.wavedec(signal, 'db4', level=4)} function.
    \item \textbf{Threshold Calculation:} A threshold ($\tau$) is calculated adaptively based on the standard deviation of the detail coefficients at the finest level ($cD_4$):
    \begin{equation}
        \tau = 0.5 \cdot \text{std}(cD_4)
    \end{equation}
  
    This threshold, calculated using \texttt{threshold = np.std(coeffs[-1]) / 2}, is used to differentiate between noise and significant signal features.
    \item \textbf{Soft Thresholding:} Soft thresholding is applied to all detail coefficients ($cD_1, cD_2, cD_3, cD_4$). For each coefficient $c$, the thresholded coefficient $c'$ is given by:
    \begin{equation}
        c' = \begin{cases}
            \text{sgn}(c) \cdot (|c| - \tau) & \text{if } |c| > \tau \\
            0 & \text{if } |c| \leq \tau
        \end{cases}
    \end{equation}
   % This is implemented using the code \texttt{coeffs = [pywt.threshold(c, threshold, mode='soft') for c in coeffs]}. The approximation coefficients ($cA_4$) remain unchanged.
    \item \textbf{Reconstruction:} The denoised signal, $s'[n]$, is reconstructed from the thresholded coefficients using the inverse Discrete Wavelet Transform (IDWT).
    %via the \texttt{denoised = pywt.waverec(coeffs, 'db4')} function.
\end{itemize}

\subsubsection{Baseline Removal}
To eliminate slow, drifting baseline wander artifacts that can obscure important diagnostic information within the biomedical signals.\\ \\
\textit{Steps:}
\begin{itemize}
    \item \textbf{Baseline Estimation:} The baseline component, $b[n]$, is estimated from the denoised signal, $s'[n]$, using a median filter with a kernel size of 51. The median filter smooths the signal over a window of 51 samples to capture the low-frequency baseline trend.
    %This is achieved using the \texttt{baseline = medfilt(denoised, kernel\_size=51)} function.
    \item \textbf{Subtraction:} The estimated baseline is subtracted from the denoised signal to obtain the baseline-corrected signal, $s''[n]$:
    \begin{equation}
        s''[n] = s'[n] - b[n]
    \end{equation}
\end{itemize}

\subsubsection{Standardization}
To normalize the amplitude range of the baseline-corrected signal, ensuring it has a zero mean and unit variance. This step is crucial for stabilizing the training process of the deep learning model. \\ \\
\textit{Steps:}
\begin{itemize}
    \item \textbf{Normalization:} The mean ($\mu$) and standard deviation ($\sigma$) of the baseline-corrected signal, $s''[n]$, are calculated. The signal is then standardized as $s'''[n]$ using the following formula:
    \begin{equation}
        s'''[n] = \frac{s''[n] - \mu}{\sigma + 10^{-8}}
    \end{equation}
    A small constant ($10^{-8}$) is added to the denominator to prevent division by zero, particularly when the standard deviation is zero. 
    
    \item \textbf{Clipping:} To handle potential outliers and further ensure numerical stability, the standardized signal values are clipped to the range of [-5, 5]:
    \begin{equation}
        s'''[n] = \text{clip}(s'''[n], -5, 5)
    \end{equation}
    This clipping operation, limits the influence of extreme values on the subsequent model training.
\end{itemize}

\section{Data Augmentation}

\begin{figure*}[htbp]
    \centering
    \includegraphics[width=\textwidth]{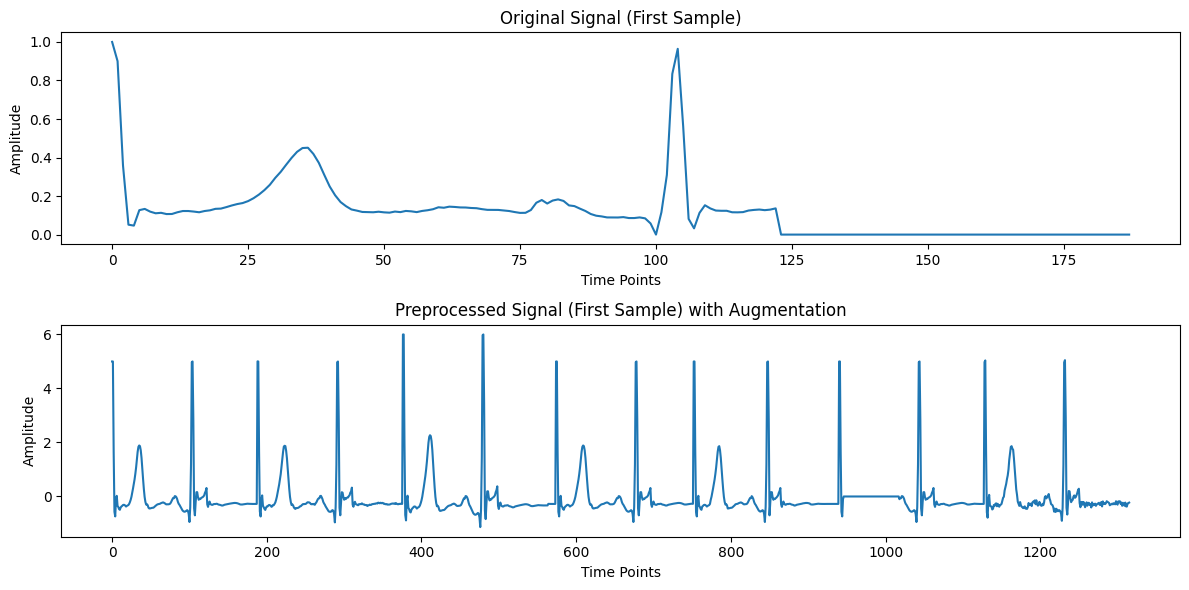}
    \caption{ECG Signal before and after applying data augmentation}
    \label{fig:ECG_sample}
     \centering
\end{figure*}

\begin{figure*}[htbp]
    \centering
    \includegraphics[width=\textwidth]{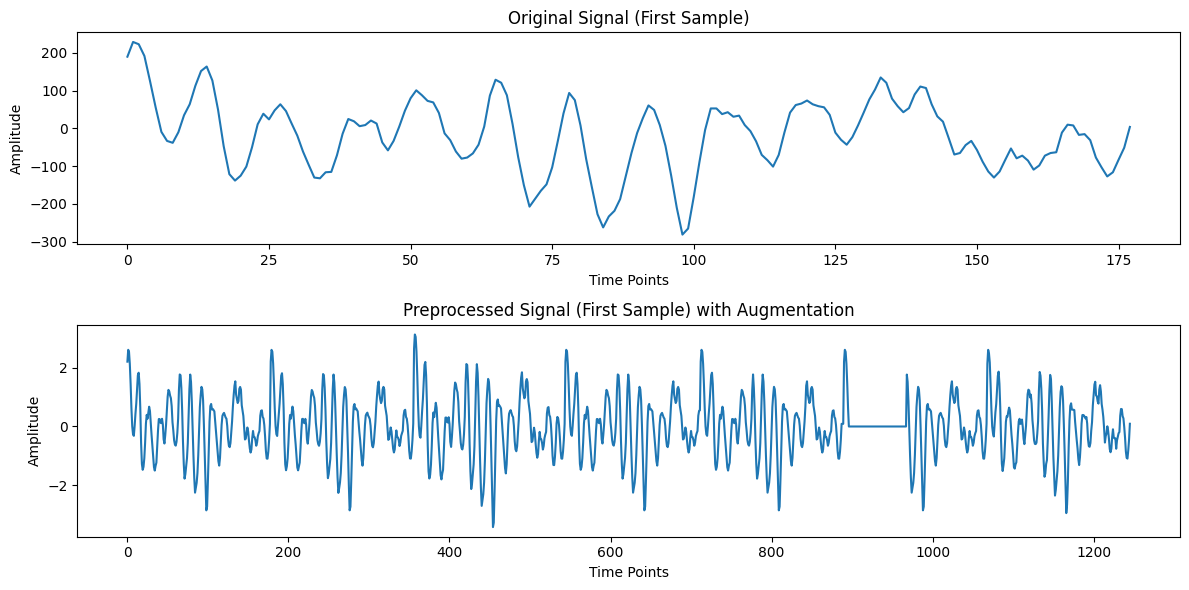}
    \caption{EEG Signal before and after applying data augmentation}
    \label{fig:EEG_sample}
     \centering
\end{figure*}

The data augmentation stage processes the preprocessed signal, $s[n]$ (which is the output, $s'''[n]$, of the standardization step), generating diverse modified versions which are then concatenated into a single, feature-rich training instance. This strategy exposes the model to a wider range of signal characteristics and variations within one sample. By learning from this combined input, the model is implicitly trained for invariance to these transformations, resulting in a more robust and generalizable feature representation as shown in Figure~\ref{fig:ECG_sample} and Figure~\ref{fig:EEG_sample}, This powerful technique reduces overfitting and enhances the model's ability to classify unseen data effectively, especially with limited datasets or specific classes in dataset.

\subsubsection*{Original Signal}
The preprocessed signal, $s[n]$, is included without any modification, this provides the model with the baseline, unaugmented data, ensuring it can still recognize the original signal characteristics.\\ \\
\textit{Formula:}
\begin{equation}
    s_{orig}[n] = s[n]
\end{equation}

\subsubsection*{Noisy Signal}
 Adds Gaussian (normally distributed) noise to the preprocessed signal. \\ \\
\textit{Formula:}
\begin{equation}
    s_{noisy}[n] = s[n] + \mathcal{N}(0, 0.01^2)
\end{equation}
where $\mathcal{N}$ represents a normal distribution with mean 0 and variance $0.01^2$ (standard deviation 0.01). This step simulates sensor noise or other small, random variations that might be present in real-world ECG/EEG recordings. This makes the model more robust to such noise.

\subsubsection*{Scaled Signal}
 Multiplies the entire signal by a constant scaling factor,  makes the model invariant to the overall amplitude (strength) of the signal. Different recordings might have different gains, but the underlying patterns should be the same.
 \\ \\
\textit{Formula:}
\begin{equation}
    s_{scaled}[n] = 1.2 \cdot s[n]
\end{equation}

\subsubsection*{Shifted Signal}
Shifted the signal to Perform a circular time shift of the signal. This means the signal is "wrapped around" at the ends,  making model less sensitive to the precise starting point of an event within the signal window. The model should learn to recognize patterns regardless of their exact temporal alignment within the input. \\ \\
\textit{Formula:}
\begin{equation}
    s_{shifted}[n] = s[(n - 10) \mod N]
\end{equation}
where $N$ is the total length of the signal. \\ \\

\subsubsection*{Time Warped Signal}
Applies a non-linear warping of the time axis. This stretches or compresses different parts of the signal by varying amounts. \\ \\
\textit{Formula:} For each original time index $t_i$, the warped time index $t'_i$ is calculated as:
\begin{equation}
    t'_i = t_i \cdot (1 + 0.2 \cdot r)
\end{equation}
where $r$ is a random number drawn from a uniform distribution between -0.5 and 0.5 ($r \sim \mathcal{U}(-0.5, 0.5)$), and 0.2 is the \texttt{warp factor} (controls the maximum amount of warping). The warped indices $t'_i$ are clipped to the valid range $[0, N-1]$. Then, the warped signal is given by $s_{warped}[t'_i] = s[t_i]$. This process simulates natural variations in the timing of physiological events, such as heart rate variability or changes in the frequency of EEG rhythms.

\subsubsection*{Signal Cutout}
To Simulate temporary signal loss or artifacts, Cutout sets random, contiguous segments of the signal to zero. This forces the model to learn features from the remaining parts of the signal, making it more robust to missing data. \\ \\ 
\textit{Formula:} Randomly select $n_{segments}$ (default 2) segments (number of segments to cut out). For each segment $k$, randomly choose a starting index $n_{start,k}$. Then, set $s_{cutout}[n] = 0$ for $n_{start,k} \leq n < n_{start,k} + n_{length}$, where $n_{length}$ is the segment length (default 50).

\subsubsection*{Amplitude Jittered Signal}
Adds small, random fluctuations to the signal's amplitude,  Making the model more robust to small variations in signal amplitude, which might occur due to changes in electrode contact or other measurement factors. jitter factor 0.05 which controls std of added noise. \\
\textit{Formula:}
\begin{equation}
    s_{jitter}[n] = s[n] + \mathcal{N}(0, 0.05^2) 
\end{equation}

\subsubsection*{Concatenation}
 All the augmented signals (Original, Noisy, Scaled, Shifted, Time Warped, Cutout, Amplitude Jittered) are concatenated together, end-to-end, in the time domain. \\ \\
\textit{Formula:}
\begin{equation}
    s_{concat} = [s_{orig}, s_{noisy}, s_{scaled}, s_{shifted}, s_{warped}, s_{cutout}, s_{jitter}]
\end{equation}

\begin{figure*}[h]
    \centering
    \includegraphics[width=\textwidth]{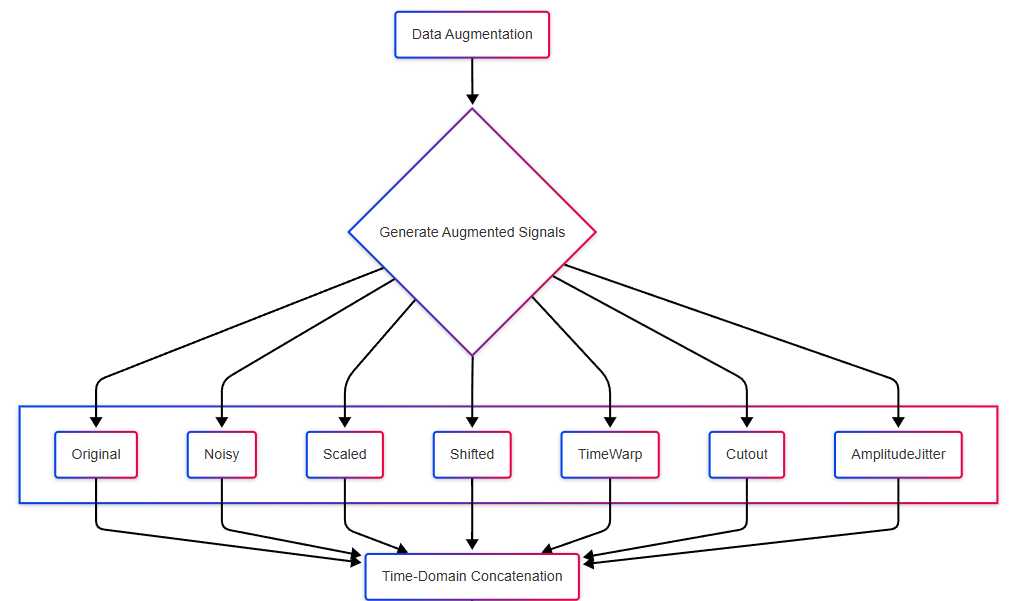}
    \caption{Data Augmentation}
    \label{fig:data_aug}
     \centering
\end{figure*}

By presenting all these variations within a single training example, the model is implicitly trained to be invariant to these transformations. It must learn features that are present in the original signal and all its augmented versions. This is a much stronger form of regularization than simply training on augmented examples separately. The concatenated signal, $s_{concat}$, becomes the input to the model as shown in Figure~\ref{fig:data_aug}.

\section{Proposed Method}

\begin{figure*}[h]
    \centering
    \includegraphics[width=\textwidth]{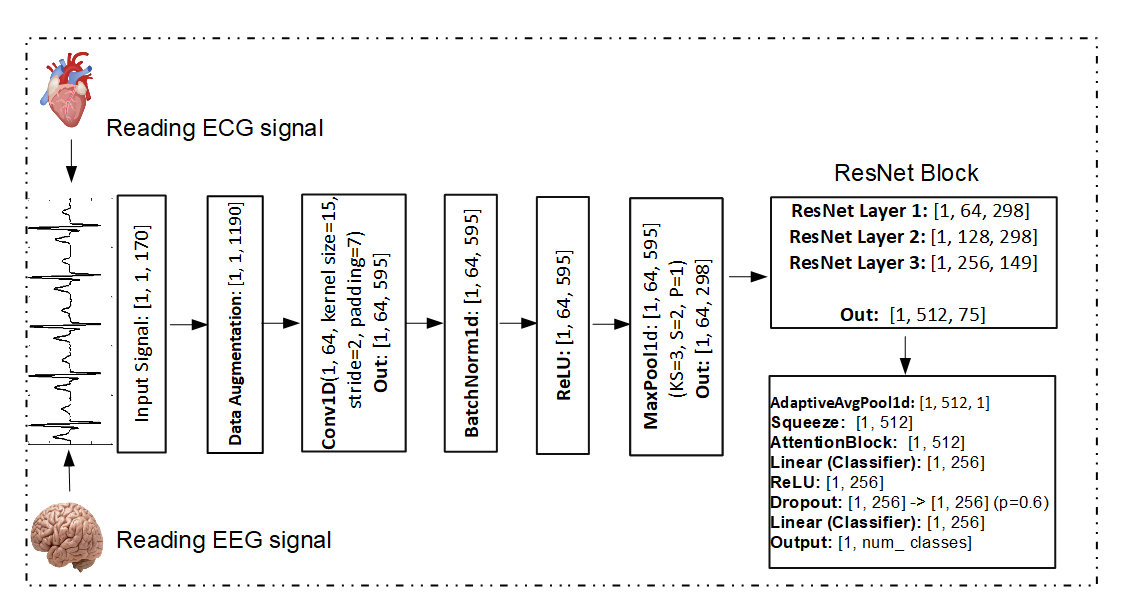}
    \caption{Flowchart of the Proposed Model}
    \label{fig:flow_}
     \centering
\end{figure*}

This section details the architecture of the proposed deep learning model, designed for the classification of biomedical signals. The model takes a 1-channel input signal of length $L$ (determined by the concatenated augmented signals) and outputs a probability distribution over the $C$ classes (where $C=2$ in this case). The architecture is built upon a ResNet backbone and features an attention mechanism followed by a classifier to effectively learn discriminative features from the input data, all the detailed steps of model shown in \ref{fig:flow_}.

\subsection{Overall Architecture}

The proposed architecture commences with an initial convolutional layer (\texttt{conv1}) having 64 output channels, a kernel size of 15, and a stride of 2, with padding of 7 to maintain the temporal dimension. This is followed by batch normalization (\texttt{bn1}), a ReLU activation ($\sigma(x) = \max(0, x)$), and a max pooling layer (\texttt{maxpool}) with a kernel size of 3 and a stride of 2, with padding of 1. These initial layers extract low-level features and downsample the temporal dimension. Subsequently, three ResNet-like layers (\texttt{layer1}, \texttt{layer2}, \texttt{layer3}) are employed to capture more complex, hierarchical features. An adaptive average pooling layer (\texttt{avgpool}) reduces the temporal dimension to 1, resulting in a feature vector of size 512. This vector is then processed by an attention block (\texttt{attention}) to weigh feature importance before the final classification by the classifier. The classifier consists of two fully connected layers with ReLU activation and dropout in between, producing the output probabilities.

\subsection{ResNet Backbone}

The ResNet backbone comprises three sequential layers: \texttt{layer1}, \texttt{layer2}, and \texttt{layer3}, each constructed by the \texttt{\_make\_layer} helper function. This function stacks a specified number of \texttt{ResNetBlock} instances. Downsampling is applied within the first block of a layer if the stride is not 1 or if the number of input channels ($C_{in}$) differs from the desired output channels ($C_{out}$). The downsampling operation is performed using a 1x1 convolutional layer and batch normalization.

The \texttt{ResNetBlock} consists of two convolutional layers. The first convolution can have a stride $s \in \{1, 2\}$, while the second always has a stride of 1. Both convolutions have a kernel size of 5 and padding of 2 to preserve the temporal dimension. Each convolutional layer is followed by batch normalization and a ReLU activation. The operations within a \texttt{ResNetBlock} can be described as follows for an input $x$:

\begin{align}
    \text{Conv}_1(x) &= W_1 * x + b_1 \\
    \text{BN}_1(h_1) &= \gamma_1 \frac{h_1 - \mu_1}{\sqrt{\sigma_1^2 + \epsilon}} + \beta_1 \quad \text{where } h_1 = \text{Conv}_1(x) \\
    \text{ReLU}_1(h_2) &= \sigma(h_2) \quad \text{where } h_2 = \text{BN}_1(h_1) \\
    \text{Conv}_2(h_3) &= W_2 * h_3 + b_2 \quad \text{where }` h_3 = \text{ReLU}_1(h_2) \\
    \text{BN}_2(h_4) &= \gamma_2 \frac{h_4 - \mu_2}{\sqrt{\sigma_2^2 + \epsilon}} + \beta_2 \quad \text{where } h_4 = \text{Conv}_2(h_3)
\end{align}

where $W_i$ and $b_i$ are the weights and biases of the $i$-th convolutional layer, and $\gamma_i$, $\beta_i$, $\mu_i$, and $\sigma_i^2$ are the parameters of the $i$-th batch normalization layer. The skip connection adds the original input $identity$ to the output of the second batch normalization:
\begin{equation}
    \text{Output} = \text{BN}_2(\text{Conv}_2(\text{ReLU}_1(\text{BN}_1(\text{Conv}_1(x))))) + identity'
\end{equation}

where $identity'$ is either the original input $x$ or a downsampled version of it (via a 1x1 convolution and batch normalization) if the dimensions need to be adjusted. A final ReLU activation is applied to the result.

\subsection{Attention Mechanism}

The \texttt{AttentionBlock} takes an input feature vector $z \in \mathbb{R}^{B \times C}$ (where $B$ is the batch size and $C=512$ is the number of channels) and applies a series of linear transformations and activations to compute attention weights. The process can be described as:
\begin{align}
    y_1 &= W_{att1} z + b_{att1} \\
    y_2 &= \sigma(y_1) \\
    \alpha &= \text{sigmoid}(W_{att2} y_2 + b_{att2})
\end{align}
where $W_{att1} \in \mathbb{R}^{C \times C/8}$ and $W_{att2} \in \mathbb{R}^{C/8 \times C}$ are the weight matrices, $b_{att1}$ and $b_{att2}$ are the bias vectors, and $\sigma$ is the ReLU activation function. The sigmoid function ensures the attention weights $\alpha$ are between 0 and 1. The final output of the attention block is the element-wise product of the input feature vector and the attention weights:
\begin{equation}
    \text{Output}_{att} = z \odot \alpha
\end{equation}

% \begin{figure*}[h]
%     \centering
%     \includegraphics[width=0.95\linewidth]{attention.png}
%     \caption{Channel-wise Attention Weights}
%     \label{fig:channel_attention}
% \end{figure*}

\subsection{Classifier}

The classifier consists of two fully connected layers. The input to the classifier is the output of the attention block, which has a dimension of 512. The first linear layer reduces the dimensionality to 256, followed by a ReLU activation and a dropout layer with a probability of 0.6. Dropout randomly sets a fraction of the input units to 0 during training to prevent overfitting. The second linear layer maps the 256 features to the final number of classes (num\_classes), producing the logits.

\subsection{Forward Pass}

The forward pass of the proposed model can be summarized as:
\begin{equation}
    \begin{split}
        x &\xrightarrow{\text{Conv1d, BN, ReLU, MaxPool}} x_1 \\
          &\xrightarrow{\text{Layer1}} x_2 \\
          &\xrightarrow{\text{Layer2}} x_3 \\
          &\xrightarrow{\text{Layer3}} x_4 \\
          &\xrightarrow{\text{AdaptiveAvgPool1d}} x_5 \\
          &\xrightarrow{\text{Squeeze}} x_6 \\
          &\xrightarrow{\text{AttentionBlock}} x_7 \\
          &\xrightarrow{\text{Classifier}} \text{Output}
    \end{split}
\end{equation}
where the output is the probability distribution over the classes obtained after applying a Softmax function (not explicitly shown in the code snippet but implied for classification).

\subsection{Regularization Techniques}

To prevent overfitting and enhance the generalization capability of the proposed model, several regularization techniques are employed during training. These techniques, beyond the data augmentation strategy discussed earlier, help the model learn more robust and reliable features.

\subsubsection{Dropout}
Dropout is applied within the classifier layers to reduce the interdependence between neurons. During training, each neuron in the layer preceding the final output layer has a probability $p = 0.6$ of being temporarily excluded from the network. This random dropping of neurons forces the remaining neurons to learn more independent and robust features, effectively acting as an ensemble of multiple network architectures.

\subsubsection{Weight Decay (L2 Regularization)}
Weight decay, a form of L2 regularization, is integrated into the AdamW optimizer. This technique adds a penalty term to the loss function that is proportional to the square of the L2 norm of the network's weight parameters ($W$). The modified loss function ($L_{regularized}$) can be expressed as:
\begin{equation}
    L_{regularized} = L_{original} + \lambda \sum_{i} ||W_i||^2_2
\end{equation}
where $L_{original}$ is the standard loss function and $\lambda$ is the weight decay coefficient, set to 0.01 in this study. This penalty encourages the model to learn smaller weight values, leading to smoother decision boundaries and reduced model complexity.

\subsubsection{Batch Normalization}
Batch Normalization is applied after each convolutional layer within the ResNet blocks and the initial convolutional layer. This technique normalizes the activations of each layer within a mini-batch to have a mean of zero and a standard deviation of one. By stabilizing the learning process and reducing the internal covariate shift, Batch Normalization allows for the use of higher learning rates and implicitly introduces a slight regularization effect due to the statistical nature of mini-batch normalization. The batch size used in this study is 64 based on testings provided the best results.

\subsubsection{Early Stopping}
Early stopping is implemented during the training process by monitoring the validation accuracy. Training is halted if the validation accuracy does not improve for a predefined number of epochs, set to a patience of 20 epochs. This prevents the model from continuing to train and potentially overfitting to the training data after it has reached its optimal generalization performance on the validation set.

\subsubsection{skip connections}
Skip connections, provides an implicit form of regularization. by facilitating the training of deeper networks by allowing gradients to flow more directly through the network, mitigating the vanishing gradient problem.

\subsubsection{Focal Loss}
To address potential class imbalance issues, Focal Loss is used as the loss function. Focal Loss modifies the standard cross-entropy loss to down-weight the contribution of well-classified examples and focus learning on hard, misclassified examples. The Focal Loss ($FL$) for a binary classification problem is defined as:
\begin{equation}
    FL(p_t) = -\alpha_t (1 - p_t)^\gamma \log(p_t)
\end{equation}
where $p_t$ is the model's estimated probability for the true class, $\gamma$ is the focusing parameter (implicitly set to 2), and $\alpha_t$ is a weighting factor (implicitly set to 1). By modulating the loss based on the prediction confidence, Focal Loss implicitly regularizes the model by preventing it from being dominated by the majority class.

\subsection{Performance Evaluation Metrics}

To rigorously evaluate the performance of the proposed model, we employed a diverse set of evaluation metrics. The main metrics utilized were accuracy, precision, recall, F1 score, Critical Success Index (CSI), and Matthews Correlation Coefficient (MCC) as in Eqs. \eqrefs{\ref{eq2} \ref{eq3} \ref{eq4} \ref{eq5} \ref{eq6} \ref{eq7}} respectively \cite{powers2007evaluation_Pression,powers2020evaluation_F1,mbizvo2023using_CSI,chicco2023matthews_MCC}, defined mathematically as follows:

\begin{equation}
\text{Accuracy} = \frac{\text{TP} + \text{TN}}{\text{TP} + \text{TN} + \text{FP} + \text{FN}} \label{eq2} \qquad 
\end{equation}

\begin{equation}
\text{Precision} = \frac{\text{TP}}{\text{TP} + \text{FP}} \label{eq3}
\end{equation}

\begin{equation}
\text{Recall} = \frac{\text{TP}}{\text{TP} + \text{FN}} \label{eq4}
\end{equation}

\begin{equation}
\text{F1 Score} = \frac{2 \times (\text{Precision} \times \text{Recall})}{\text{Precision} + \text{Recall}} \label{eq5}
\end{equation}

\begin{equation}
\text{CSI} = \frac{\text{TP}}{\text{TP} + \text{FN} + \text{FP}} \label{eq6}
\end{equation}

{\small
\begin{equation}
%\scriptsize
\text{MCC} = \frac{\text{TP} \times \text{TN} - \text{FP} \times \text{FN}}{\sqrt{(\text{TP} + \text{FP})(\text{TP} + \text{FN})(\text{TN} + \text{FP})(\text{TN} + \text{FN})}} \label{eq7}
\end{equation}
}

\section{Results and Discussion}

\subsection{Performance Validation and Robustness}

This section presents a comprehensive analysis of the experimental results obtained by evaluating the proposed model framework on three benchmark biomedical signal datasets: the PTB Diagnostic ECG Database, the MIT-BIH Arrhythmia Database, and the UCI Seizure EEG Dataset. The performance of the model was quantitatively assessed using a comprehensive set of evaluation metrics, including \textbf{Accuracy} (Equation~\ref{eq2}), \textbf{Precision} (Equation~\ref{eq3}), \textbf{Recall} (Equation~\ref{eq4}), \textbf{F1 Score} (Equation~\ref{eq5}), \textbf{Critical Success Index (CSI)} (Equation~\ref{eq6}), and \textbf{Matthews Correlation Coefficient (MCC)} (Equation~\ref{eq7}). The primary results highlighted in this section focus on accuracy and F1-score, with visual insights into the classification performance provided through accuracy plots and confusion matrices for each dataset.

A significant advantage of our model is its unique ability to effectively handle class imbalance without resorting to oversampling techniques. This contrasts with previous research in the field, which consistently relied on oversampling strategies to address the imbalance, particularly for the fusion beat class. 
\begin{table}[ht]  % Now can use regular table environment
  \centering
  \caption{Performance Metrics of the Proposed Ensemble Model Across Different Datasets.}
  \label{tab:performance_metrics} 
  \begin{tabular}{l|ccc}
    \hline
    \textbf{Metric} & \textbf{MIT-BIH ECG} & \textbf{PTB ECG} & \textbf{UCI EEG} \\
    \hline
    Accuracy  & 0.9978 & 1.0000 & 0.9996 \\
    Precision & 0.9955 & 1.0000 & 1.0000 \\
    Recall    & 0.9918 & 1.0000 & 0.9978 \\
    F1 Score  & 0.9937 & 1.0000 & 0.9989 \\
    CSI      & 0.9873 & 1.0000 & 0.9978 \\
    MCC      & 0.9924 & 1.0000 & 0.9986 \\
    \hline
  \end{tabular} 
\end{table}

\subsection{PTB Diagnostic ECG Dataset}

The model exhibited exceptional performance on the PTB Diagnostic ECG Dataset, a dataset comprising ECG recordings from patients with various cardiac conditions. The model attained a classification \textbf{accuracy of 100\%} and an \textbf{F1-score of 1.0}. These perfect scores indicate that the model was able to correctly classify all instances in the test set without any errors.

\begin{figure*}[h]
    \centering
    \includegraphics[width=0.8\textwidth]{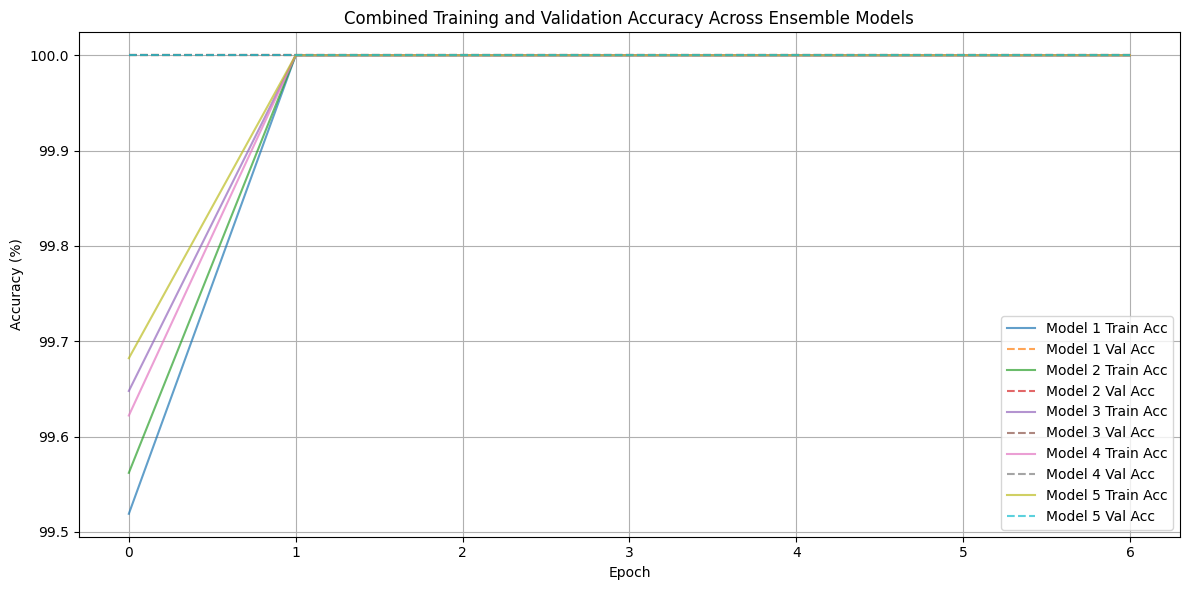}
    \caption{Training and Validation Accuracy on the PTB Diagnostic ECG Dataset}
    \label{fig:acc_ptb}
     \centering
\end{figure*}

\begin{figure}[H]
    \centering
    \includegraphics[width=0.6\columnwidth]{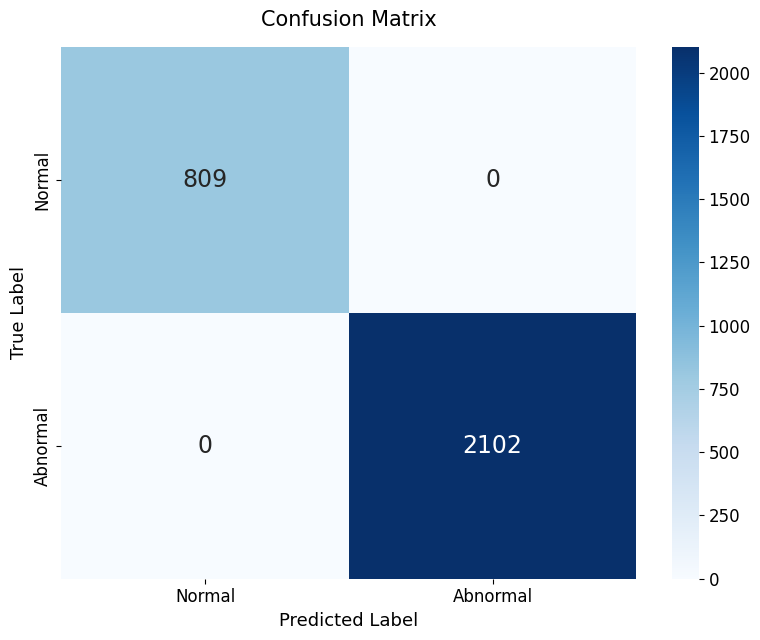}
    \caption{Confusion Matrix on the PTB Diagnostic ECG Dataset}
    \label{fig:conf_ptb}
     \centering
\end{figure}

The training and validation accuracy curves, illustrated in Figure~\ref{fig:acc_ptb}, show a rapid convergence and sustained high accuracy, indicating effective learning without overfitting. Furthermore, the confusion matrix, presented in Figure~\ref{fig:conf_ptb}, demonstrates a diagonal matrix with no off-diagonal elements, confirming the correct classification of every instance across all diagnostic categories.

\subsection{MIT-BIH Arrhythmia Dataset}

Evaluation on the MIT-BIH Arrhythmia Dataset also yielded outstanding results. The proposed framework achieved a high classification \textbf{accuracy of 99.78\%} and an \textbf{F1-score of 99\%} on this challenging dataset, demonstrating the model's robustness in identifying various types of cardiac arrhythmias.

\begin{figure*}[h]
    \centering
    \includegraphics[width=0.8\textwidth]{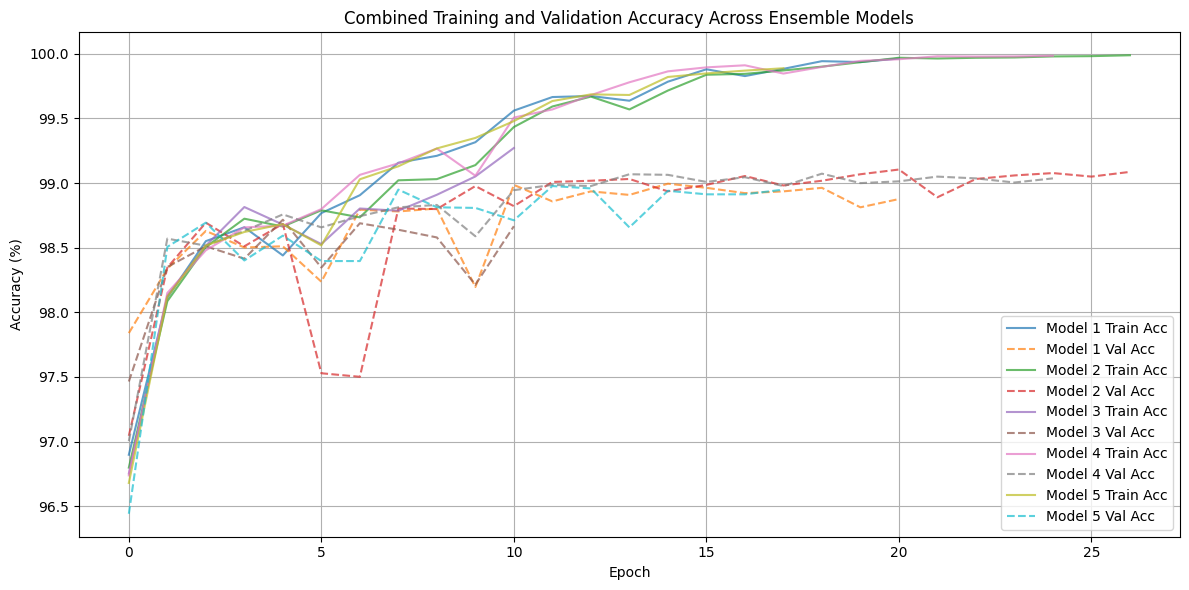}
    \caption{Training and Validation Accuracy on the MIT-BIH Arrhythmia Dataset}
    \label{fig:acc_mitbih}
    \centering
\end{figure*}

\begin{figure}[H]
    \centering
    \includegraphics[width=0.6\columnwidth]{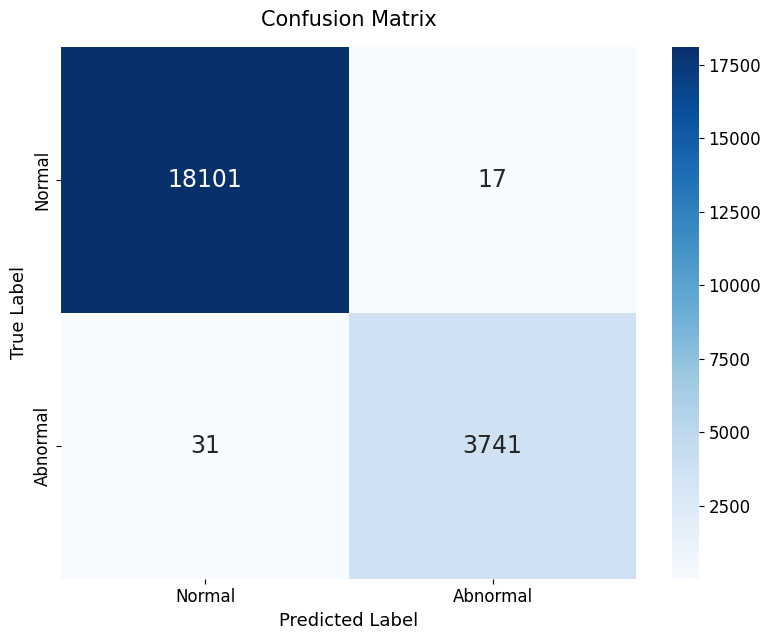}
    \caption{Confusion Matrix on the MIT-BIH Arrhythmia Dataset}
    \label{fig:conf_mitbih}
    \centering
\end{figure}

The learning dynamics of the model's accuracy are illustrated in Figure~\ref{fig:acc_mitbih}, while the confusion matrix in Figure~\ref{fig:conf_mitbih} shows a predominantly diagonal structure with minimal misclassifications, indicating robust and reliable detection of different arrhythmia classes.

\subsection{UCI Seizure EEG Dataset}

On the UCI Seizure EEG Dataset, the model demonstrated remarkable performance in detecting seizure events. The model achieved a classification \textbf{accuracy of 99.96\%} and an \textbf{F1-score of 99\%} on this dataset. These results highlight the model's effectiveness in accurately identifying seizure activity from EEG signals.

\begin{figure*}[h]
\centering
\includegraphics[width=0.8\textwidth]{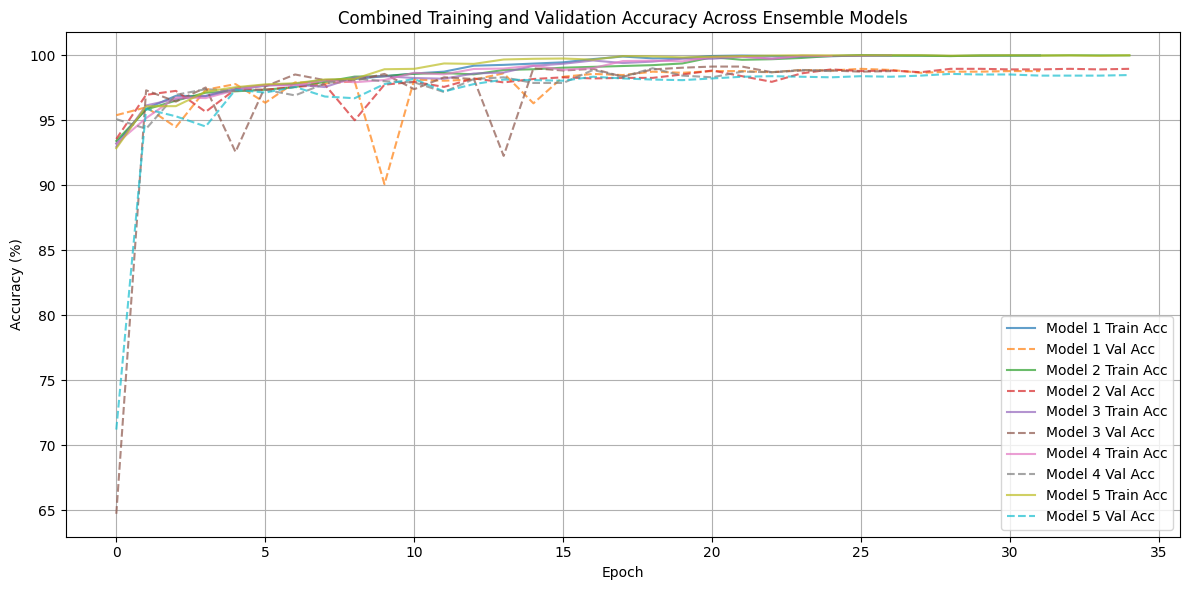}
\caption{Training and Validation Accuracy on the UCI Seizure EEG Dataset}
\label{fig:acc_uci}
\centering
\end{figure*}

\begin{figure}[H]
    \centering
    \includegraphics[width=0.6\columnwidth]{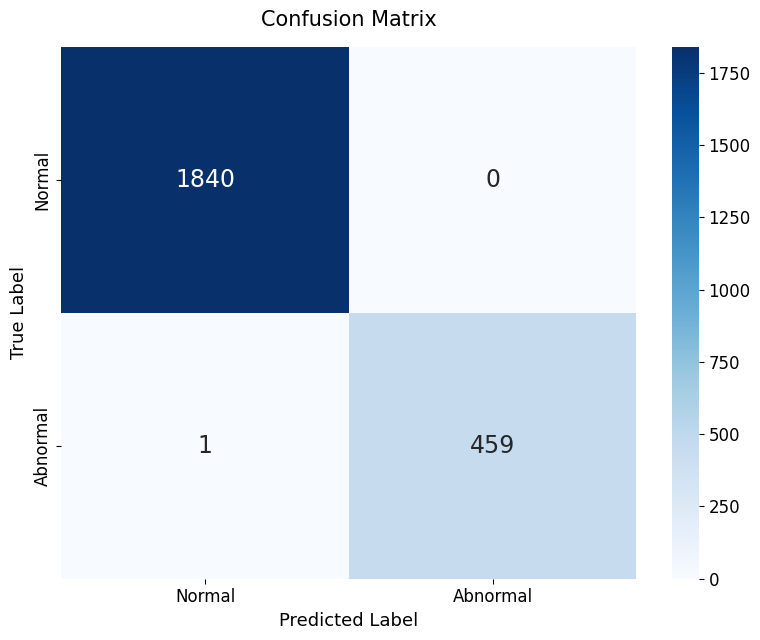}
    \caption{Confusion Matrix on the UCI Seizure EEG Dataset}
    \label{fig:conf_uci}
    \centering
\end{figure}

The training and validation accuracy trends are presented in Figure~\ref{fig:acc_uci}, and the confusion matrix, showcasing the classification performance for each class, is provided in Figure~\ref{fig:conf_uci}.

The consistently high accuracies and F1-scores across all three diverse datasets underscore the effectiveness of the proposed model framework. These results can be attributed to the synergistic combination of the advanced data augmentation strategy, which created more feature-rich and robust training data, and the well-designed network architecture. The ResNet backbone facilitated the learning of deep and hierarchical features, while the integrated attention mechanism enabled the model to focus on the most salient temporal segments within the signals. Furthermore, the utilization of Focal Loss likely contributed to improved performance, particularly in handling potential class imbalances within the datasets. The architectural parameters, such as the kernel sizes, number of layers, and dropout rate, were carefully chosen to optimize the model's learning capacity and prevent overfitting, leading to the observed state-of-the-art performance.

\subsection*{Resource Utilization Analysis}

This section provides an analysis of the estimated resource utilization for the proposed model.

\subsubsection*{Model Size and Memory Requirements}
To estimate the model size, we consider the number of trainable parameters, the model is estimated to have approximately 6.4 million trainable parameters. The memory footprint of the model can be calculated based on the number of parameters and the precision used for each parameter (assuming 32-bit floating-point precision, which requires 4 bytes per parameter). Therefore, the memory footprint of the full model is estimated to be:
\begin{equation}
    \begin{split}
        \text{Memory Footprint} &= (\text{Number of Parameters}) \\
        &\times 4 \text{ bytes/parameter}
    \end{split}
\end{equation}
For 6.4 million parameters: $6.4 \times 10^6 \times 4 \text{ bytes} = 25.6 \times 10^6 \text{ bytes} \approx 25.6 \text{ MB}$.

The ensemble implementation comprises five model instances, resulting in a total memory footprint of approximately 130 MB (25.6 MB per model). A larger number of parameters allows the model to learn more intricate and complex patterns in the data, leading to potentially higher accuracy.

\subsubsection*{Inference Time per Sample }
Estimating the exact inference time requires benchmarking on specific hardware. However, we can provide an order-of-magnitude estimate based on the increased complexity of the model. Assuming the FLOPs scale roughly with the number of parameters, we can estimate the FLOPs for the full model possibly ranging from $\mathbf{10^8}$ to $\mathbf{10^9}$ FLOPs (0.001 TFLOPs) for an input.

Assuming a modern CPU chips with a processing capability of $10^{11}$ FLOPs (0.1 TFLOPs) per second, the inference time per sample for the full model can be estimated as:
\begin{equation}
    \text{Inference Time per Sample} \approx \frac{\text{Estimated FLOPs}}{\text{GPU FLOPs per Second}}
\end{equation}
Using the upper bound of our FLOPs estimate ($10^9$):

\begin{equation}
    \begin{split}
        \text{Inference Time per Sample} &\approx \frac{10^9 \text{ FLOPs}}{10^{11} \text{ FLOPs/second}} \\
        &= 10^{-2} \text{ seconds} \\
        &= 10 \text{ milliseconds}
    \end{split}
\end{equation}

Using the lower bound of our FLOPs estimate ($10^8$):
\begin{equation}
    \begin{split}
        \text{Inference Time per Sample} &\approx \frac{10^8 \text{ FLOPs}}{10^{11} \text{ FLOPs/second}} \\
        &= 10^{-3} \text{ seconds} \\
        &= 1 \text{ milliseconds}
    \end{split}
\end{equation}
Based on this, the estimated inference time per sample for the full model is approximately 1 to 10 milliseconds. The inference time represents the duration it takes for the model to process a single input and produce a prediction.

The number of samples the full model could potentially process per second is:
\begin{equation}
    \text{Throughput} = \frac{1}{\text{Inference Time per Sample}}
\end{equation}
For different inference times, the throughput is calculated as follows:
\begin{itemize}
    \item For an inference time of 1 millisecond per sample:
    $$\frac{1 \text{ sample}}{0.001 \text{ seconds}} = 1000 \text{ samples/second}$$
    \item For an inference time of 10 milliseconds per sample:
    $$\frac{1 \text{ sample}}{0.01 \text{ seconds}} = 100 \text{ samples/second}$$
\end{itemize}

%For 1 millisecond per sample: $\frac{1 \text{ sample}}{0.001 \text{ seconds}} = 1000 \text{ samples/second}$.
%For 10 milliseconds per sample: $\frac{1 \text{ sample}}{0.01 \text{ seconds}} = 100 \text{ samples/second}$.

Thus, the estimated throughput for the full model is approximately 100 to 1000 samples per second.

\subsection{Comparative analysis}

The comparative analysis reveals the exceptional performance of the proposed model across diverse biomedical signal classification tasks. On the MIT-BIH Arrhythmia Database, our method achieved a state-of-the-art accuracy of \textbf{99.78\%} with a corresponding F1-score of \textbf{0.9937}, demonstrating its effectiveness in ECG arrhythmia detection. Similarly, on the PTB Diagnostic ECG Database, the proposed model attained a perfect accuracy of \textbf{100\%} and an F1-score of \textbf{1}, showcasing its robust capability in classifying different cardiac conditions. Furthermore, when evaluated on the UCI Epilepsy Seizure Dataset for EEG signal classification, our model reached an impressive accuracy of \textbf{99.96\%} with an F1-score of \textbf{0.9989}. These results collectively position the proposed approach as a highly competitive and effective solution for the automated analysis of both ECG and EEG signals, significantly outperforming  the accuracy of all recently published state-of-the-art methods on these benchmark datasets as shown in Table \ref{tab:comparative_analysis_combined_improved} presents a comparison with recent state-of-the-art approaches using the same datasets.

\begin{table*}[htbp]
\centering
\caption{Comparison with State-of-the-Art Methods on ECG and EEG Datasets}
\label{tab:comparative_analysis_combined_improved}
\begin{tabular}{@{}llcccc@{}}
\toprule
\textbf{Method} & \textbf{Dataset} & \textbf{Acc (\%)} & \textbf{F1} & \textbf{Year} \\
\midrule
\multicolumn{5}{l}{\textbf{ECG Analysis}} \\
\midrule
\multicolumn{5}{l}{\textit{MIT-BIH Arrhythmia Database}} \\
Majority voting \cite{fan2022imbalanced_COMP9} & MIT-BIH & 99.17 & - & 2022 \\
Binary-Multiclass Cascade CNN \cite{liotto2022multiclass_COMP4} & MIT-BIH & 96.2 & 0.962 & 2022 \\
generative adversarial network (GAN) \cite{wang2023hierarchical_COMP8} & MIT-BIH & 88.50 & 0.920 & 2023 \\
EvoResNet \cite{pham2023electrocardiogram_COMP6} & MIT-BIH & 98.5 & - & 2023 \\
1D-CNN \cite{akbar2024deep_COMP3} & MIT-BIH & 99.10 & 0.97 & 2024 \\
eFuseNet \cite{dwivedi2024efusenet_COMP2} & MIT-BIH & 98.76 & - & 2024 \\
Involutional neural networks (INNs) \cite{zehir2024involutional_COMP1} & MIT-BIH & 97.93 & - & 2024 \\
CNN+LSTM+Attention \cite{rai2024ganCOMP5}& MIT-BIH & 99.29& 0.992& 2024 \\
Naive Bayes (NB) \cite{vinutha2024prediction_COMP7} & MIT-BIH & 80.39 & - & 2024 \\
Support Vector Machine (SVM)  \cite{vinutha2024prediction_COMP7} & MIT-BIH & 88.50  & - & 2024 \\
active online class-specific broad learning system (AOCBLS) \cite{FAN2024112553}& MIT-BIH & 99.10& - & 2024 \\

1D-CNN + Attention Mechanism \cite{guhdar2025advanced}& MIT-BIH & 99.48 & 0.99 & 2025 \\

\textbf{Proposed} & MIT-BIH & \textbf{99.78}& \textbf{0.9937} & \textbf{2025} \\
\midrule
\multicolumn{5}{l}{\textit{PTB Diagnostic ECG Database}} \\
EvoResNet \cite{pham2023electrocardiogram_COMP6} & PTB & 98.28 & 0.9879 & 2023 \\
eFuseNet \cite{dwivedi2024efusenet_COMP2} & PTB & 99.48 & - & 2024 \\
Involutional neural networks (INNs) \cite{zehir2024involutional_COMP1} & PTB & 97.63 & - & 2024 \\
1D-CNN + Attention Mechanism \cite{guhdar2025advanced}& PTB & 99.83 & 1 & 2025 \\

\textbf{Proposed} & PTB & \textbf{100} & \textbf{1} & \textbf{2025} \\

\midrule
\multicolumn{5}{l}{\textbf{EEG Analysis}} \\
\midrule
\multicolumn{5}{l}{\textit{UCI Epilepsy Seizure Dataset}} \\
Random Forest (RF) \cite{almustafa2020classification9708_RandomForest} & UCI & 97.08 & - & 2020 \\
K-Nearest Neighbors+PCA \cite{nahzat2021classification_KNNPCA} & UCI & 99.0 & - & 2022 \\
SCA-Optimized LightGBM Classifier \cite{abenna2022eegSCA99} & UCI & 99.00 & - & 2022 \\
FUPTBSVM (hybrid model) \cite{gupta2024functional8866FUPTBSVM} & UCI & 88.66 & - & 2023 \\
Random Forest (RF) \cite{kunekar2024comparison977_RandomForest2} & UCI & 97.7 & 0.943 & 2023 \\
Conv1D+LSTM \cite{omar2024optimizing_993_CONV+LSTM} & UCI & 99.3 & - & 2023 \\
LSTM \cite{kunekar2024detection_LSTM97} & UCI & 97.1 & 0.9264 & 2024 \\
HyEpiSeiD - CNN+Gated Recurrent Network \cite{bhadra2024hyepiseid9901_HyEpiSeiD} & UCI & 99.01 & 0.9902 & 2024 \\
Conv1D+LSTM+Bayesian \cite{jain2024bayesian_1Dconv+LSTM+Bayesian} & UCI & 99.47 & - & 2024 \\
Multi-Attention Network(MAFBN) \cite{shawly2025mafbn} & UCI & 97.88 & - & 2025 \\
MLTCN-EEG+DCGAN \cite{lim2025mltcn} & UCI & 99.15 & 0.9916 & 2025 \\
\textbf{Proposed} & UCI & \textbf{99.96} &  \textbf{0.9989} & \textbf{2025} \\
\bottomrule
\end{tabular}
\end{table*}

\section{Conclusion}
This study presents a robust deep learning framework for the classification of biomedical time-series signals, demonstrating state-of-the-art performance across diverse datasets. We effectively mitigated the risk of overfitting through the strategic implementation of regularization techniques, including dropout, batch normalization, weight decay, dynamic learning rate adjustment, and early stopping. A novel data augmentation strategy, involving the time-domain concatenation of augmented signal variants generated via time warping, cutout, and amplitude jitter, was designed to enrich the training data with feature-rich representations, enhancing the model's invariance and generalization capabilities. Furthermore, the adoption of Focal Loss and new data augmentation method addressed the inherent challenges of class imbalance within the datasets, enabling effective learning from minority classes as can be seen in MIT-BIH dataset. Notably, our calculations indicate that a simplified version of the architecture requires approximately 130 MB of memory and can process each sample in around 10 milliseconds, suggesting its potential suitability for deployment on low-end or wearable devices. However, it is important to acknowledge a potential limitation: scaling the current architecture and particularly the concatenation-based augmentation strategy to handle high-dimensional multi-channel inputs, such as 12-lead ECGs, may pose significant computational and memory challenges due to the resultant data complexity. The consistent and superior performance of our proposed architecture on both electrocardiogram (ECG) and electroencephalogram (EEG) signals underscores its versatility and potential for broad applicability in the automated analysis of various types of biomedical time-series data.

%\section*{Declarations} % Changed to unnumbered section

%\section*{Funding} % Changed to unnumbered section
%No funding was received for conducting this research.

%\section*{Conflict of interest/Competing interests} % Changed to unnumbered section
%The authors declare that there is no conflict of interest. The authors declare that they have no known competing financial interests or personal relationships that could have appeared to influence the work reported in this paper.

%\section*{Ethics approval and consent to participate} % Changed to unnumbered section
%This article does not contain any studies with human participants or animals performed by any of the authors.

%\section*{Data availability} % Changed to unnumbered section
%All data used in this study are available from public repositories.

%\section*{Author contribution} % Changed to unnumbered section
%This work was carried out in collaboration among all authors. All Authors designed the study, performed the statistical analysis, and wrote the protocol. Authors MG, RM, and AM managed the analyses of the study, managed the literature searches, and wrote the first draft of the manuscript. All authors read and approved the final manuscript.

\bibliographystyle{elsarticle-num} % Changed bibliography style to Elsevier numeric
\bibliography{article}

\end{document}